\documentclass[fleqn,usenatbib]{mnras}
\usepackage{newtxtext,newtxmath}

\usepackage[T1]{fontenc}
\usepackage{ae,aecompl}

\usepackage{graphicx}	
\usepackage{amsmath}	
\usepackage{amssymb}	
\usepackage{times}

\newcommand{\pc}                        {\,{\rm pc}}

\newcommand{\pkpc}                      {\,{\rm pkpc}}

\newcommand{\cMpc}                      {\,{\rm cMpc}}
\newcommand{\Msun}                    {\,{\rm M}_\odot}

\newcommand{\Msunyr}                 {\,{\rm M}_\odot\,{\rm yr}^{-1}}

\newcommand{\K}                          {\,{\rm K}}

\title[The gas fractions of $\sim L^\star$ galaxy haloes]{The gas fractions of dark matter haloes hosting simulated $\sim L^\star$ galaxies are governed by the feedback history of their black holes}

\author[J. J. Davies et al.]{Jonathan  J. Davies,$^{1}$\thanks{E-mail: j.j.davies@2016.ljmu.ac.uk} 
Robert A. Crain,$^{1}$ 
Ian G. McCarthy,$^{1}$ 
Benjamin D. Oppenheimer,$^{2}$ 
\newauthor 
Joop Schaye,$^{3}$
Matthieu Schaller$^{3}$
and Stuart McAlpine$^{4}$ 
\\
$^{1}$Astrophysics Research Institute, Liverpool John Moores University, 146 Brownlow Hill, Liverpool L3 5RF\\
$^{2}$CASA, Department of Astrophysical and Planetary Sciences, University of Colorado, 389 UCB, Boulder, CO 80309, USA\\
$^{3}$Leiden Observatory, Leiden University, PO Box 9513, NL-2300 RA Leiden,
the Netherlands\\
$^{4}$Department of Physics, University of Helsinki, Gustaf H\"allstr\"amin katu 2a, 00560 Helsinki, Finland
}

\date{Accepted XXX. Received YYY; in original form ZZZ}

\pubyear{2018}

\hypersetup{draft} 
\begin{document}
\label{firstpage}
\pagerange{\pageref{firstpage}--\pageref{lastpage}}
\maketitle

\begin{abstract}
We examine the origin of scatter in the relationship between the gas fraction and mass of dark matter haloes hosting present-day $\sim L^\star$ central galaxies in the EAGLE simulations. The scatter is uncorrelated with the accretion rate of the central galaxy's black hole (BH), but correlates strongly and negatively with the BH's mass, implicating differences in the expulsion of gas by active galactic nucleus feedback, throughout the assembly of the halo, as the main cause of scatter. Haloes whose central galaxies host undermassive BHs also tend to retain a higher gas fraction, and exhibit elevated star formation rates (SFRs). Diversity in the mass of central BHs stems primarily from diversity in the dark matter halo binding energy, as these quantities are strongly and positively correlated at fixed halo mass, such that $\sim L^\star$ galaxies hosted by haloes that are more (less) tightly-bound develop central BHs that are more (less) massive than is typical for their halo mass. Variations in the halo gas fraction at fixed halo mass are reflected in both the soft X-ray luminosity and thermal Sunyaev-Zel'dovich flux, suggesting that the prediction of a strong coupling between the properties of galaxies and their halo gas fractions can be tested with measurements of these diagnostics for galaxies with diverse SFRs but similar halo masses.
\end{abstract}




\newpage
\section{Introduction}

Analysis of the latest measurements of the primordial anisotropies exhibited by the cosmic microwave background constrain the cosmic baryon fraction, $\Omega_{\rm b} / \Omega_0 \simeq 0.15$, with sub-percent precision \citep{planck18}. Panoramic galaxy surveys indicate that approximately 5 percent (by mass) of these baryons are in the form of stars and stellar remnants \citep[see e.g.][]{balogh01,liwhite09,baldry12}, implying that the vast majority remains in gaseous form, leading to the notion of a `missing baryons' problem.

In the absence of ejective feedback processes, one would expect the majority of the missing (i.e. non-stellar) baryons to be diffuse gas associated with collapsed haloes \citep[e.g.][]{crain07}. However, feedback mechanisms that eject gas from the interstellar medium, and potentially also the haloes of galaxies, appear to be a necessary component of galaxy formation models as a means to regulate galaxy growth \citep[primarily via star formation feedback, e.g.][]{oppenheimer08,oppenheimer10,crain09,puchwein13,schaye10}, and to eventually quench their star formation and regulate the cooling of intragroup/intracluster gas via feedback associated with the growth of black holes (BHs) \citep[e.g.][]{croton06,bower06,sijacki06,fabjan10,mccarthy10}. Estimates of the mass of circumgalactic gas based on absorption features in quasar sightlines or X-ray emission fall significantly short of the cosmic baryon fraction \citep[e.g.][]{bregman07,shull12,werk13}, and only in the massive haloes of rich galaxy clusters, with dynamical masses $\sim 10^{15}\Msun$, is the baryon fraction observed to converge to the cosmic average \citep[e.g.][]{allen02,lin04,gonzalez13}.

Halo gas, often termed the circumgalactic medium (CGM), is the interface between galaxies and the intergalactic medium (IGM), and acts as reservoir of both the fuel for ongoing star-formation, and of the heavy elements whose synthesis accompanies this process. Numerical simulations of galaxy evolution demonstrate that the structure, temperature, element abundances and ionisation state of the CGM are markedly sensitive to the implementation of feedback processes \citep[e.g.][]{vandevoort12,hummels13,ford16}. Such processes remain poorly-constrained by observations and their physical efficiencies cannot (yet) be predicted from first principles, meaning that ab initio prediction of the relationship between the gas fraction and total mass of haloes is not yet feasible.

Here we examine the influence of galaxy properties on the present-day gas fractions of haloes in the EAGLE simulations of galaxy formation. EAGLE adopts the pragmatic approach of calibrating feedback efficiencies to ensure the reproduction of key properties of the galaxy population, such as their stellar and central BH masses and the sizes of disc galaxies\footnote{Section 2 of \cite{schaye15} motivates this approach in detail.}, and has been shown to reproduce a diverse range of observable properties of the galaxy population \citep[e.g][]{furlong15,furlong17,trayford15,trayford17,segers16,crain17} and intergalactic gas, as probed by X-ray emission \citep[e.g.][]{schaye15} and absorption features in quasar sightlines \citep[e.g.][]{oppenheimer16,rahmati15,rahmati16,turner17}. The suite is therefore well suited to the study of the co-evolution of galaxies and their gaseous environments. 

In Section \ref{sec:methods} we briefly describe the simulations and our techniques for identifying and characterising galaxies and their haloes. In Section \ref{sec:origin_of_scatter} we present the scaling relation between the halo gas fraction and halo mass, and examine the origin of scatter about it, whilst in Section \ref{sec:obs} we investigate means by which these predictions of the simulations can be confronted with observational measurements. We summarise and discuss our findings in Section \ref{sec:summary}. In Appendix \ref{app:obs} we briefly compare the X-ray and thermal Sunyaev-Zel'dovich fluxes of our simulated haloes with current observational measurements. Throughout, we adopt the convention of prefixing units of length with `c' and `p' to denote, respectively, comoving and proper scales, e.g. cMpc for comoving megaparsecs. 

\section{Methods}
\label{sec:methods}

\subsection{Numerical simulations}
\label{sec:sim}

EAGLE \citep[Evolution and Assembly of GaLaxies and their Environments,][]{schaye15,crain15} is a suite of cosmological hydrodynamical simulations that follow the formation and evolution of galaxies\footnote{The public release of EAGLE data is described by \citet{mcalpine16}.}. EAGLE adopts a $\Lambda$CDM cosmogony described by the parameters advocated by the \cite{planck14}, namely $\Omega_0 = 0.307$, $\Omega_{\rm b} =
0.04825$, $\Omega_\Lambda= 0.693$, $\sigma_8 = 0.8288$, $n_{\rm s} = 0.9611$, $h = 0.6777$ and $Y = 0.248$. The simulations were evolved using a version of the smoothed particle hydrodynamics (SPH) and Tree-PM gravity solver GADGET-3 \citep[last described by][]{springel05}, incorporating several modifications to the hydrodynamics scheme. These include an implementation of the pressure-entropy SPH formulation of \citet{hopkins13}, the time-step limiter of \citet{durier12}, and switches for artificial viscosity and artificial conduction of the forms proposed by \citet{cullen10} and \citet{price10}, respectively.

The EAGLE software also includes a number of subgrid treatments of processes operating below the numerical resolution of the simulation, including radiative cooling \citep{wiersma09}; star formation \citep{schaye08}; stellar evolution and mass loss \citep{wiersma09b}; BH formation, growth and mergers \citep{springeldimatteohernquist05,rosasguevara15,schaye15}; and feedback associated with the formation of stars \citep[`stellar feedback',][]{dallavecchiaschaye12} and the growth of BHs \citep[`AGN feedback',][]{boothschaye09}. The efficiency of stellar feedback was calibrated to reproduce the present-day stellar masses of galaxies and the sizes of galaxy discs, whilst the efficiency of AGN feedback was calibrated to reproduce the present-day scaling relation between the stellar mass of galaxies and that of their central BH \citep[for further details see][]{crain15}. The gaseous properties of galaxies and their haloes were not considered during the calibration and may be considered predictions of the simulations.

We analyse four simulations from the EAGLE suite. We focus primarily on the simulation with the largest volume and greatest particle number, Ref-L100N1504, which evolves with the EAGLE Reference model a periodic cube of side $L = 100\,{\rm cMpc}$, populated with $N=1504^3$ collisionless dark matter particles with mass $9.70\times 10^6\Msun$ and an (initially) equal number of baryonic particles with mass $1.81\times 10^6\Msun$. In order to compute the intrinsic binding energy of haloes in this simulation, i.e. that which emerges in the absence of the dissipative physics of galaxy formation (see Section \ref{sec:sample}), we also analyse a simulation starting from identical initial conditions but considering only collisionless gravitational dynamics, DMONLY-L100N1504. We briefly examine NOAGN-L050N0752, a simulation following a smaller $L = 50\,{\rm cMpc}$ cubic volume at the same resolution, using a variation of the Reference model in which AGN feedback is disabled. To ensure that comparisons with this simulation are made on an equal footing we use Ref-L050N0752, a simulation of the same $L = 50\,{\rm cMpc}$ volume using the EAGLE Reference model. In all cases a Plummer-equivalent gravitational softening length of $\epsilon_{\rm com} = 2.66\,{\rm ckpc}$ was used, limited to a maximum proper length of $\epsilon_{\rm prop} = 0.7\,{\rm pkpc}$. 

\subsection{Characterising haloes and galaxies}
\label{sec:sample}

Haloes are identified by applying the friends-of-friends algorithm to the dark matter particle distribution, with a linking length of 0.2 times the mean interparticle separation. Gas, stars and BHs are associated with the FoF group, if any, of their nearest dark matter particle. Bound substructures are subsequently identified within haloes using the SUBFIND algorithm \citep{springel01,dolag09}. We consider present-day haloes with $M_{200} > 10^{11.5}\Msun$, with each halo thus being resolved by at least $\sim 10^5$ particles. The typical present-day stellar mass of central galaxies hosted by haloes with $M_{200} \simeq 10^{11.5}\Msun$ is $M_\star \simeq 10^{9.5}\Msun$; as shown by \citet{schaye15}, present-day galaxies in EAGLE with at least this mass exhibit a passive fraction that broadly agrees with observational measurements. 

We compute the spherical overdensity mass \citep[$M_{200}$,][]{laceycole94} of each halo about its most-bound particle, such that the mean density enclosed within the sphere of radius $r_{200}$ is $200$ times the critical density, $\rho_{\rm c}$. More generally, halo properties are computed by aggregating the properties of all particles of the relevant type that reside within an appropriate aperture. We compute the inner halo binding energy by summing the binding energies of all particles within $r_{2500}$ that comprise each halo's counterpart in the DMONLY-L100N1504 simulation. In common with a number of other studies based on the analysis of EAGLE \citep[e.g.][]{schallerconcs,matthee18}, we use this `intrinsic' binding energy, $E_{\rm DMO}^{2500}$, in order to eliminate the influence of dissipative baryonic processes. The dissipation of baryons in, and their ejection from, the progenitors of haloes throughout their formation and assembly can markedly influence their structure, potentially masking or exaggerating the influence of the intrinsic properties of the haloes. We pair haloes with their counterparts using the bijective particle matching algorithm described by \citet{schallerconcs}, which successfully pairs 3411 of the 3543 haloes (96 percent) satisfying $M_{200} > 10^{11.5}\Msun$ in Ref-L100N1504. Unpaired haloes are discarded, thus ensuring the same sample of haloes is used throughout. 

Following \citet{schaye15}, we compute the properties of central galaxies by aggregating the properties of the relevant particles that reside within $30\pkpc$ of the halo centre. We equate the BH mass of galaxies, $M_{\rm BH}$, to the mass of their most-massive BH particle, which is almost exclusively coincident with the halo centre. 

\section{The origin of scatter in halo gas fractions}
\label{sec:origin_of_scatter}

\begin{figure}
\includegraphics[width=\columnwidth]{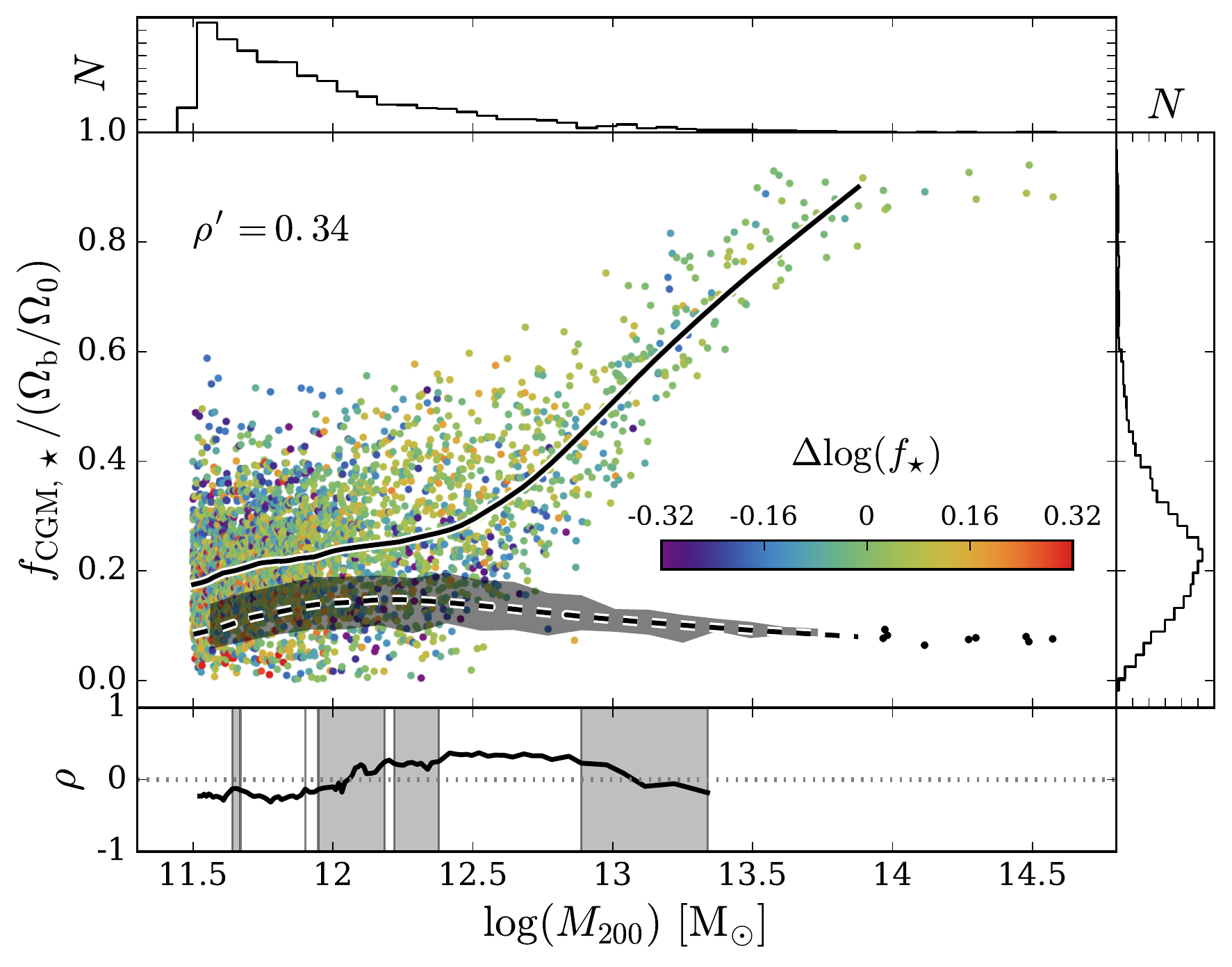} 
\caption{Present-day halo gas fractions, $f_{\rm CGM}$, as a function of halo mass, $M_{200}$. Histograms of $M_{200}$ and $f_{\rm CGM}$ are shown above and to the right of the main panel, respectively. The solid curve denotes the running median, whilst the dashed curve denotes the running median of the total stellar fraction of the halo, $f_\star$. The shaded region shows the $10^{\rm th}-90^{\rm th}$ percentile scatter of $f_\star$. Symbols are coloured by the residuals of the relationship between $f_\star$ and $M_{200}$, $\Delta \log_{10} f_\star$. The lower panel shows the running value of the Spearman rank coefficient, $\rho$, for the $\Delta f_{\rm CGM} - \Delta \log_{10} f_\star$ relation. Shading denotes where the recovered correlation is not significant ($p > 0.01$).}
\label{fig:Ref100_mstar}
\end{figure}

In Fig. \ref{fig:Ref100_mstar} we show, as a fraction of their mass $M_{200}$, the circumgalactic gas fraction, $f_{\rm CGM} \equiv M_{\rm gas}(r<r_{200})/M_{200}$, normalised by the cosmic baryon fraction, of present-day haloes in the Ref-L100N1504 simulation\footnote{The baryon and stellar fractions of EAGLE haloes were presented and discussed by \citet{schallerconcs}.}. This definition includes the contribution of interstellar gas, but this is in general a small fraction of the halo gas mass. The solid black line shows the running median of the gas fraction, $\tilde{f}_{\rm CGM}(M_{200})$, computed via the locally-weighted scatterplot smoothing method \citep[LOWESS, e.g.][]{cleveland79}. There is considerable scatter in $f_{\rm CGM}$ in relatively low-mass haloes, which declines for $M_{200} \gtrsim 10^{13}\Msun$.

For reference, the running median of the total stellar mass fraction of the halo, $\tilde{f}_\star(M_{200})$ is also shown as a dashed black curve, where $f_\star \equiv M_\star(r<r_{200})/M_{200}$. The shaded region about this curve denotes the $10^{\rm th}-90^{\rm th}$ percentile scatter of $f_\star$. The LOWESS curves are plotted within the interval for which there are at least 10 measurements at both higher and lower $M_{200}$; in poorly-sampled high-mass bins, halo stellar fractions are plotted as individual black dots. Histograms of $M_{200}$ and $f_{\rm CGM}$ are shown above and to the right, respectively, of the main panel. Gas fractions transition from $\simeq 0.3 \Omega_{\rm b}/\Omega_0$ below $M_{200} \simeq 10^{12.5}\Msun$, rising steadily towards $\simeq 0.9 \Omega_{\rm b}/\Omega_0$ at $M_{200} \simeq 10^{14}\Msun$, beyond which the trend flattens. This interval therefore represents a transition regime between EAGLE's relatively gas-poor, low-mass haloes and their gas-rich, high-mass counterparts. Present-day $\sim L^\star$ galaxies, with stellar mass similar to that of the Milky Way \citep[$M_\star \simeq 6\times 10^{10}\Msun$, e.g.][]{blandhawthorn16} are thought to be hosted by haloes with mass $M_{200}\simeq 10^{12.5}\Msun$ \citep[e.g.][]{moster13}
 
Symbols are coloured by the residuals of the relationship between the stellar mass fraction and the halo mass, i.e for the $i^{\rm th}$ halo, $\Delta \log_{10} f_{\star, i} = \log_{10} f_{\star, i} - \log_{10} \tilde{f}_\star(M_{200, i})$. Haloes denoted by red (blue) points therefore have a greater (lower) stellar mass fraction than is typical for their halo mass. Inspection of the symbol colours indicates that $\Delta f_{\rm CGM}$ and $\Delta \log_{10} f_\star$ are not strongly correlated at any mass scale. We quantify the strength of the correlation with the Spearman rank correlation coefficient, $\rho$. Since the correlations can in principle exhibit a strong dependence on halo mass, we compute `running' correlation coefficients from halo-mass ordered sub-samples. For bins whose median halo mass exhibits $M_{200}<10^{12}\Msun$, we use samples of 300 haloes with starting ranks separated by 50 haloes (i.e. 1-300, 51-350, 101-400 etc), otherwise we obtain superior sampling of the high-mass range with bins of 100 haloes separated by 25 haloes\footnote{The resulting coefficients are not strongly sensitive to these choices.}. This diagnostic is shown in the bottom panel of Fig. \ref{fig:Ref100_mstar}. In this and subsequent figures we shade regions where the $p$-value exceeds $0.01$, corresponding to < $2.3\sigma$ confidence, to highlight where the recovered correlation is not significant. The correlation coefficient for the $\Delta f_{\rm CGM} - \Delta \log_{10} f_\star$ is relatively low ($|\rho|\lesssim 0.3$) at all halo masses, and is recovered with low confidence for much of the halo mass range. The coefficient for the 106 haloes in a $0.1\,{\rm dex}$ window about $M_{200}= 10^{12.5}\Msun$ can however be recovered with significance, and its value, which for reference we denote $\rho^\prime$, is $0.34$. This constitutes a weak-to-moderate correlation for a narrow range in $M_{200}$, but more broadly it is evident that the diversity of the gas fractions exhibited by present-day $\sim L^\star$ haloes does not emerge primarily as a consequence of some haloes converting more of their gas into stars than others. We note however that, as one approaches the `closed box' regime of massive haloes with near-unity baryon fractions, one should expect a correlation between the scatter in $f_{\rm CGM}(M_{200})$ and $f_\star(M_{200})$ to emerge \citep{farahi18}.

\begin{figure*}
\includegraphics[width=\textwidth]{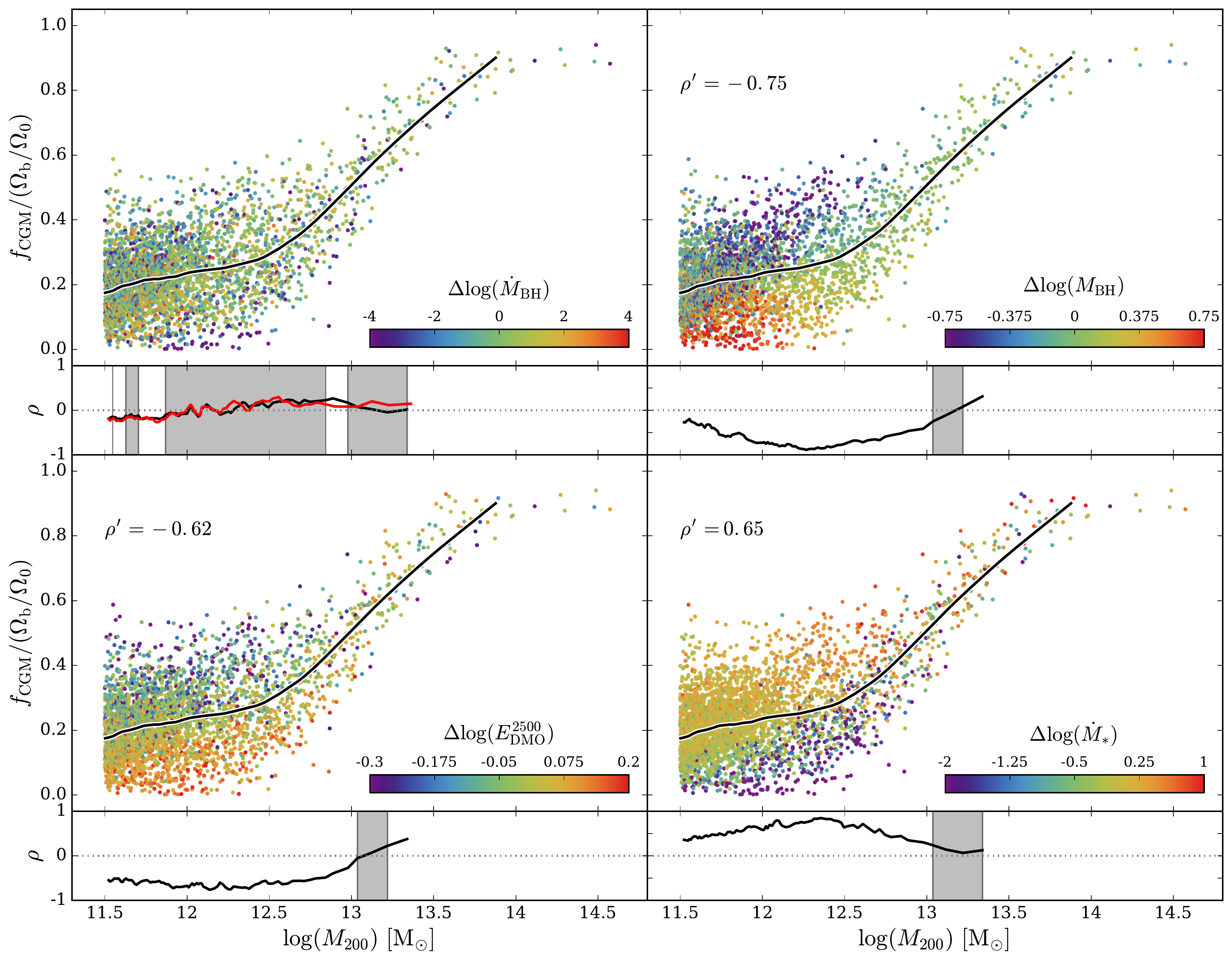} 
\caption{Present-day halo gas fractions, $f_{\rm CGM}$, as a function of halo mass, $M_{200}$, with the solid curve denoting the running median. In each of the main panels, the symbols are coloured by residuals about the relationships of various properties as a function of halo mass: the accretion rate of the central BH ($\dot{M}_{\rm BH}$, \textit{top left}), the mass of the central BH ($M_{\rm BH}$, \textit{top right}), the intrinsic binding energy of the inner halo ($E_{\rm DMO}^{2500}$, \textit{bottom-left}), and the star formation rate of the central galaxy ($\dot{M}_\star$, \textit{bottom-right}). Beneath the main panels we show the running value of the Spearman rank correlation coefficient for the relationships between these residuals and $\Delta f_{\rm CGM}$, the residuals about the $f_{\rm CGM}-M_{200}$ relation.  The red curve in the upper-left plot corresponds to the running $\rho$ recovered when smoothing the BH accretion rate over a $100\,{\rm Myr}$ window. Shading denotes regions for which the correlation is recovered at low significance ($p>0.01$). In the cases where correlations are significant, we quote the Spearman coefficient, $\rho^\prime$, of the correlation computed for haloes within a 0.1dex window about $\log_{10} M_{200} [\Msun] = 12.5$.}
\label{fig:Ref100_megaplot}
\end{figure*}

The rest-mass energy liberated throughout the growth of central BHs is comparable to the gravitational binding energy of the halo \citep[][see also \citealt{oppenheimer18}]{silkrees98}, and is thus expected to foster the expulsion of gas from galaxies and their haloes. It is reasonable to surmise then that AGN feedback should have a significant influence on the gas fraction of haloes \citep[e.g.][]{puchwein08,bower08,mccarthy10,mccarthy11,bocquet16,pillepich18}. In Fig. \ref{fig:Ref100_megaplot} we again show the present-day $f_{\rm CGM} - M_{200}$ relation of Ref-L100N1504, with the symbols in the top row coloured by the residuals about the median relation between the characteristics of central BHs and halo mass. In the top-left panel they are coloured by the residuals about the running median of the BH accretion rate as a function of halo mass, $\Delta \log_{10} \dot{M}_{\rm BH}$. As in the case of the $\Delta \log_{10} f_\star$ correlation shown in Fig. \ref{fig:Ref100_mstar}, the Spearman rank correlation coefficient is relatively low ($|\rho| < 0.3$) at all halo masses, and is recovered with low confidence for much of the sampled mass range, including at $M_{200}=10^{12.5}\Msun$, hence we are unable to quote a significant value of $\rho^\prime$ for this diagnostic. Scatter in halo gas fractions is therefore not strongly correlated with scatter in the BH accretion rate at any $M_{200}$. \citet{mcalpine17} recently showed that the accretion rate of BHs in EAGLE can vary by orders of magnitude on very short timescales ($\lesssim 10^5\,{\rm yr}$), so we have repeated this test after time-averaging the BH accretion rate over the preceding $100\,{\rm Myr}$. We find similar results with this definition of $\dot{M}_{\rm BH}$, as is evident from its running value of $\rho$, shown as a red curve.

Prior analyses of the EAGLE simulations \citep{bower17,mcalpine18} have revealed that the development of a hot ($T \gtrsim 10^6\K$), quasi-hydrostatic CGM in haloes with mass $M_{200} \gtrsim 10^{12}\Msun$ inhibits the buoyant transport away from the galaxy of gas ejected from the interstellar medium (ISM) in stellar feedback-driven outflows. The resulting build-up of gas triggers non-linear growth of the BH, which accretes rapidly until the feedback associated with its growth becomes the dominant means of regulating the inflow of gas onto the galaxy. \citet{mccarthy11} argue that the expulsion of gas from the progenitors of group- and cluster-scale haloes, which accompanies this onset of BH feedback, occurs primarily at early cosmic epochs ($1\lesssim z \lesssim 3$) when their central BHs accreted most of their mass. We therefore colour the symbols of the top-right panel of Fig. \ref{fig:Ref100_megaplot} by the residuals about the running median of the BH mass as a function of halo mass, $\Delta \log_{10} M_{\rm BH}$. In this case, the colouring reveals a striking negative correlation between the gas fraction (at fixed halo mass) and the BH mass, such that haloes whose central galaxies host atypically-massive central BHs exhibit systematically low gas fractions, and vice versa. The visual impression is corroborated by the Spearman rank correlation coefficient, which is significant and negative for all $10^{11.5} < M_{200} \lesssim 10^{13}\Msun$. The coefficient for haloes with $M_{200} \simeq 10^{12.5}\Msun$ is very strong, $\rho^\prime=-0.75$. These results indicate that the halo gas fractions of $\sim L^\star$ galaxies are regulated primarily by the evolutionary stage of their central BHs.

We turn next to the origin of the diversity of BH masses at fixed $M_{200}$. This question was explored by \citet{boothschaye10,boothschaye11} using the OWLS simulations \citep{schaye10}, who concluded that BH mass is governed primarily by the binding energy of the inner halo. We therefore colour the symbols of the bottom-left panel of Fig. \ref{fig:Ref100_megaplot} by the residuals about the running median of the binding energy of the halo as a function of halo mass. As motivated in Section \ref{sec:sample}, we use the intrinsic binding energy, $E_{\rm DMO}^{2500}$, i.e. that recovered from each halo's counterpart in the DMONLY simulation measured within $r_{2500}$. This eliminates the effects of dissipative physical processes, which could potentially mask or exaggerate any correlations induced by the intrinsic binding energy of the halo. The colouring reveals a striking negative correlation between the scatter in $f_{\rm CGM}$ and that of the binding energy of the inner halo\footnote{Using $E_{\rm DMO}^{500}$ or $E_{\rm DMO}^{200}$ instead yields similar results.}. The Spearman rank correlation coefficient is significant and negative for all $10^{11.5} < M_{200} \lesssim 10^{13}\Msun$, and exhibits a broad, strong minimum (characterised by $\rho < -0.5$ recovered at $p < 0.01$) for $M_{200} \lesssim 10^{12.5}\Msun$. The coefficient for haloes with $M_{200} \simeq 10^{12.5}\Msun$ is strong, with $\rho^\prime=-0.62$. At fixed mass, more tightly-bound haloes require more energy to unbind gas from the inner halo, so their central BHs must grow to be more massive, reaching a higher peak luminosity, and thus they ultimately eject a greater fraction of the halo gas beyond $r_{200}$\footnote{This effect underpins the importance of using $E_{\rm DMO}^{2500}$ rather than $E_{\rm Ref}^{2500}$. In general the dissipation of gas into stars results in $E_{\rm Ref}^{2500} > E_{\rm DMO}^{2500}$, but crucially the fractional difference in the two measures at fixed halo mass anti-correlates strongly with $E_{\rm DMO}^{2500}$, because the formation of more massive BHs at the centres of haloes with greater $E_{\rm DMO}^{2500}$ enables them to eject a greater fraction of their gas. The use of $E_{\rm Ref}^{2500}$ therefore partially masks the underlying correlation between the inner binding energy and $f_{\rm CGM}$.}. More tightly-bound haloes at fixed mass also tend to be those with a higher concentration and an earlier formation time \citep[e.g.][]{navarro04}; indeed \citet{boothschaye10,boothschaye11} found that halo concentration correlates with $M_{\rm BH}$ at fixed $M_{200}$. We have therefore examined the relationships between the scatter in gas fractions at fixed halo mass and the scatter in the \citet[][`NFW']{nfw} concentration and the halo assembly lookback time \citep[computed as per][]{qu17} of each halo's counterpart in the DMONLY-L100N1504 simulation, and recover negative correlations that are again significant, albeit slightly weaker than is the case for $E_{\rm DMO}^{2500}$.

The binding energy of the halo in the DMONLY simulation is effectively encoded within the phase-space configuration of the initial conditions, and residuals of $f_{\rm CGM}$ correlate with similar strength, but over a wider range in halo mass, to those of $E_{\rm DMO}^{2500}$ than with those of $M_{\rm BH}$. Scatter in $E_{\rm DMO}^{2500}$ at fixed $M_{200}$ might reasonably then be considered as the fundamental cosmological origin of the scatter in $f_{\rm CGM}$ at fixed $M_{200}$. However, the influence of the binding energy is physically `transmitted' to the gas fraction by ejective feedback. The necessity of this conduit can be demonstrated using the NOAGN-L050N0752 EAGLE simulation, in which AGN feedback is disabled. Examination of the relationship between $\Delta f_{\rm CGM}$ and $\Delta \log_{10} E_{\rm DMO}^{2500}$ in this simulation reveals a significantly weaker correlation than in the Reference simulation, which is driven largely by the lowest-mass haloes in the sample. The origin of the correlation here differs with respect to the Reference simulation; haloes with greater binding energy (at fixed mass) exhibit lower gas fractions in the NOAGN simulation because the positive correlation of the $\Delta f_\star$ - $\Delta \log_{10} E_{\rm DMO}^{2500}$ relation is much stronger than in the Reference simulation. The halo gas fraction is therefore depleted because of the condensation of gas into stars rather than its ejection by feedback. 

\begin{figure}
\includegraphics[width=\columnwidth]{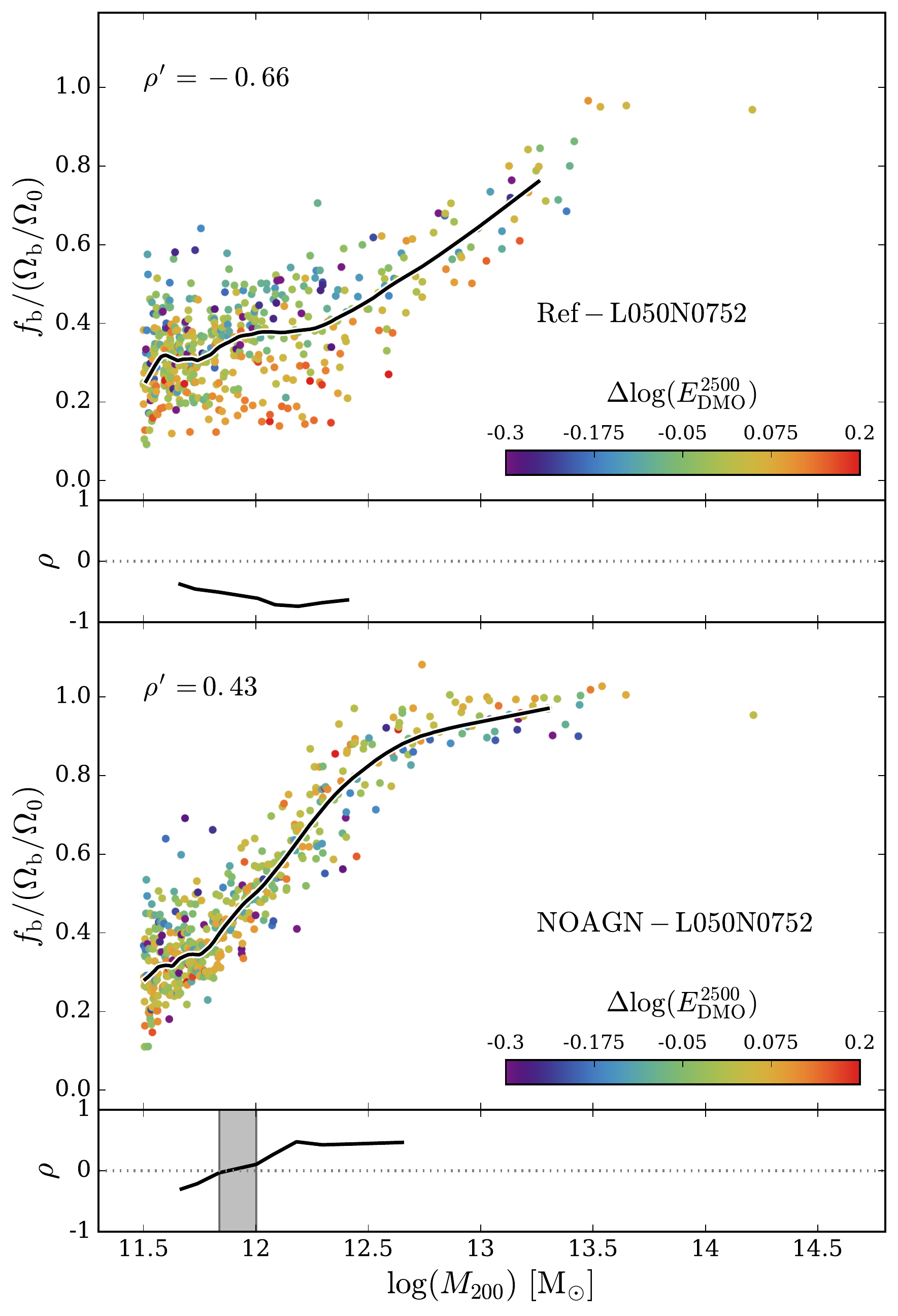} 
\caption{Present-day halo baryon fractions, $f_{\rm b}$, as a function of halo mass, $M_{200}$, in the Ref-L050N0752 (\textit{top}) and NOAGN-L050N0752 (\textit{bottom}) simulations. In each case the solid curve denotes the running median, and the symbols are coloured by the residuals about the running median of the $E_{\rm DMO}^{2500}-M_{200}$ relation, $\Delta E_{\rm DMO}^{2500}$. Sub-panels show the running value of the Spearman rank correlation coefficient, $\rho$, for the relationships between these residuals and $\Delta f_{\rm b}$, the residuals about the $f_{\rm b}-M_{200}$ relation. Shading denotes regions for which the correlation is recovered at low significance ($p>0.01$).}
\label{fig:noagn}
\end{figure}

The transmission of the influence of halo binding energy via AGN feedback can be more concisely demonstrated by examination of the halo baryon fraction, $f_{\rm b} \equiv [M_{\rm gas}(r<r_{200})+M_\star(r<r_{200})]/M_{200}$. In Fig. \ref{fig:noagn} we show the present-day $f_{\rm b}-M_{200}$ relations for Ref-L050N0752 (top) and NOAGN-L050N0752 (bottom), and colour the symbols by $\Delta \log_{10} E_{\rm DMO}^{2500}$. Using the halo baryon fraction rather than the halo gas fraction takes account of the aforementioned additional condensation of gas into stars in the NOAGN simulation. In the absence of AGN feedback, haloes with $M_{200} \gtrsim 10^{12}\Msun$ retain a significantly greater fraction of their baryons, as stellar feedback is unable to expel gas from massive haloes. In Ref-L050N0752, the $\Delta f_{\rm b} - \Delta \log_{10} E_{\rm DMO}^{2500}$ correlation for haloes with $M_{200} = 10^{11.5}-10^{13}\Msun$ is strong, significant and negative for $M_{200} > 10^{11.8}\Msun$, with $\rho^\prime = -0.66$\footnote{We compute $\rho^\prime$ for $L=50\cMpc$ simulations within the larger window of $0.5\,{\rm dex}$ around $M_{200}=10^{12.5}\Msun$, to maintain adequate sampling.}. In NOAGN-L050N0752, the correlation is again mildly negative for $M_{200} \lesssim 10^{12}\Msun$, but becomes moderately positive for more massive haloes, with $\rho^\prime = 0.43$. Therefore, in the absence of AGN feedback, the intrinsic binding energy of the inner region of haloes has a markedly different influence on the baryon fractions for $M_{200} \gtrsim 10^{12}\Msun$. This notwithstanding, we reiterate that residuals about the $f_{\rm CGM}-M_{200}$ relation in the Reference model correlate strongly with those about the $E_{\rm DMO}^{2500}-M_{200}$ relation over a broader range in halo mass than with those about the $M_{\rm BH}-M_{200}$ relation, indicating that the binding energy may also influence halo gas fractions via other mechanisms, such as formation time.

It is interesting to note that the correlations between $\Delta f_{\rm CGM}$ and each of $\Delta \log_{10} M_{\rm BH}$ and $\Delta \log_{10} E_{\rm DMO}^{2500}$ change sign for $M_{200} \gtrsim 10^{13}\Msun$, and become positive. The reversal of these correlations is likely a consequence of the declining efficiency of gas expulsion by AGN feedback in the most-massive haloes; as feedback becomes unable to eject gas from the assembling halo, a greater central binding energy only serves to inhibit gas expulsion.

\citet{matthee18} recently demonstrated that the star formation histories of EAGLE galaxies hosted by haloes with earlier formation times (at fixed halo mass) are systematically shifted to earlier cosmic epochs. A consequence of this effect is that these galaxies exhibit systematically lower present-day star formation rates (SFRs, $\dot{M}_\star$). Since haloes with early formation times generally exhibit greater central binding energies, one might expect that scatter about the $f_{\rm CGM} - M_{200}$ relation will correlate with the SFR. Returning briefly to Fig. \ref{fig:Ref100_megaplot}, we show in the bottom-right panel the $f_{\rm CGM} - M_{200}$ relation with symbols coloured by the residuals about the running median of the SFR as a function of halo mass, $\Delta \log_{10} \dot{M}_\star$, which indeed reveals a striking positive correlation between the residuals about the running medians of the $f_{\rm CGM} - M_{200}$ and $\dot{M}_\star - M_{200}$ relations. The running coefficient of the Spearman rank correlation is positive for all halo masses, and for $M_{200} \lesssim 10^{13}\Msun$ it resembles the inverse of that of the $\Delta f_{\rm CGM} - \Delta \log_{10} M_{\rm BH}$ correlation. The coefficient for haloes with $M_{200} \simeq 10^{12.5}\Msun$ is $\rho^\prime = 0.65$, highlighting the strength of this correlation. More gas-rich haloes at fixed mass, besides exhibiting relatively undermassive BHs, therefore also exhibit an elevated SFR. 

A close connection between the SFR of galaxies and their interstellar gas content is well established by observations \citep[e.g.][]{kennicutt98}, a finding whose reproduction and interpretation continues to attract considerable analytic and numerical effort \citep[e.g.][]{thompson05,krumholz07,semenov16,orr18}. To our knowledge, however, a correlation between the gas fractions of haloes (of similar mass) and the SFR of their central galaxies has not been demonstrated previously. The correlation of both the SFR and the BH mass with the halo gas fraction at fixed halo mass is helpful from the perspective of scrutinizing the predictions advanced here, as the SFR can be inferred from a diverse range of photometric and spectroscopic diagnostics; we turn to this scrutiny in the next section. 

The correlation is also of intrinsic interest. Surveys of galaxies and associated absorption systems have revealed a positive correlation between the SFR of galaxies and the column density of the absorbers \citep[e.g.][]{chen10,lan14,rubin18}, and a popular interpretation of this correlation is that the absorption column densities are enhanced by outflows driven by stellar feedback. Analysis of  EAGLE suggests that the correlation is not causal, but is rather a consequence of the negative correlation of both the halo gas fraction, and the SFR, with the halo binding energy at fixed halo mass. The first correlation is a consequence of more tightly-bound haloes requiring more massive BHs to unbind gas from their inner regions, whilst the second is due to more tightly-bound haloes collapsing earlier, shifting the growth of their central galaxy and BH (and the associated expulsion of their halo gas) to earlier times. This interpretation is likely sensitive to the details of the phenomenological implementation of the relevant physical processes in the EAGLE model, and we defer further detailed exploration of this sequence of events to a follow-up study.

\section{Testing via complementary observables}
\label{sec:obs}

\begin{figure*}
\includegraphics[width=\textwidth]{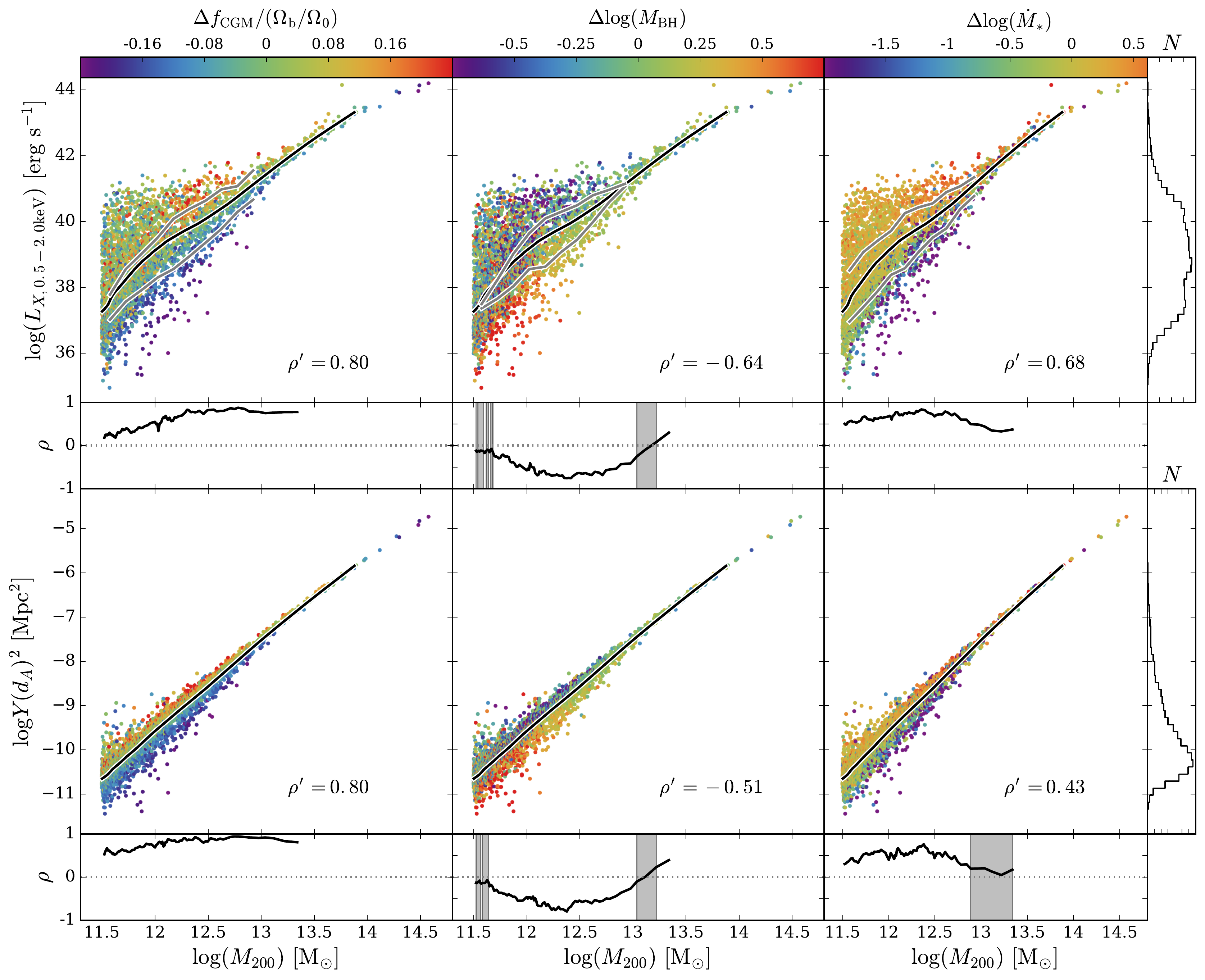} 
\caption{Present-day diffuse soft X-ray luminosity ($L_{\rm X}$, \textit{top}) and thermal Sunyaev-Zel'dovich effect flux ($Y d^2_{\rm A}$, \textit{bottom}), as a function of halo mass, $M_{200}$. Histograms of $L_{\rm X}$ and $Y d^2_{\rm A}$ are shown to the right of the main panel. The solid curve denotes the running median of each quantity. In each column the symbols are coloured by the residuals about the running median of $f_{\rm CGM}$ (\textit{left}), $M_{\rm BH}$ (\textit{centre}) and $\dot{M}_\star$ (\textit{right}) as a function of $M_{200}$. Beneath the main panels we show the running value of the Spearman rank correlation coefficient for the relationships between these residuals and $\Delta f_{\rm CGM}$, the residuals about the $f_{\rm CGM}-M_{200}$ relation. Shading denotes regions for which the correlation is recovered at low significance ($p>0.01$), whilst $\rho^\prime$ denotes the value of the Spearman rank correlation coefficient computed for haloes with mass $M_{200} \simeq 10^{12.5}\Msun$. The grey curves on the upper row show the median of the $L_{\rm X}-M_{200}$ relations of the halo sub-samples comprising the upper and lower quartiles of the diagnostic used for the symbol colouring.}
\label{fig:Ref100_Lx_tSZ}
\end{figure*}

The influence of central BH mass on the scatter of halo gas fractions at fixed halo mass is an unambiguous prediction of the EAGLE simulations. However, it is not one that is trivial to confront with observations, since one requires measurements of $f_{\rm CGM}$ and $M_{200}$, and dynamical measurements of BH masses, for a large sample of galaxies. We therefore briefly explore in this section whether it might be possible to test the predictions of Section \ref{sec:origin_of_scatter} with complementary observational diagnostics provided by extant or forthcoming facilities.  

We require diagnostics for which scatter about their running median as a function of halo mass correlates with scatter about the $f_{\rm CGM}-M_{200}$ relation. We first consider the diffuse, soft ($0.5 \lesssim E_{\rm X} \lesssim 2.0\,{\rm keV}$) X-ray luminosity of the hot ($T \gtrsim 10^6\,\K$), collisionally-ionized component of halo gas. Characterisation of the properties of the gas in haloes less massive than those of galaxy groups ($M_{500}\sim 10^{13}\Msun$, $kT \sim 1\,{\rm keV}$) remains challenging, with the extended hot CGM of only a handful of galaxies having been convincingly detected and characterised beyond the optical envelope of the galaxy \citep[e.g.][]{dai12,bogdan13b,bogdan17,li16,li17}, but forthcoming and proposed X-ray observatories such as \textit{Athena} and \textit{Lynx} promise to make such detections more commonplace. Moreover, stacking low spatial resolution \textit{ROSAT} All-Sky Survey X-ray maps about the coordinates of local, optically-selected galaxies has proven an effective means of characterising the relationship between galaxies and their gas content \citep{anderson15,wang16}. We therefore compute the soft X-ray luminosity of each halo by coupling the physical properties of its constituent gas particles (i.e. those within $r_{200}$) to the Astrophysical Plasma Emission Code \citep[APEC,][]{smith01}, using the techniques described by \citet[ see also \citealt{crain10,crain13}]{mccarthy17}. 

Stacking about the coordinates of optically-selected galaxies has also recently been used to characterise the hot gas content of haloes via measurement of the thermal Sunyaev-Zel'dovich (tSZ) effect, the inverse Compton scattering of cosmic microwave background (CMB) photons by energetic electrons within the hot, ionized CGM \citep[e.g.][]{planck13,greco15}. The tSZ `flux' can be defined as the Compton-$y$ parameter integrated over the solid angle of the halo and is thus proportional to the total energy of the hot gas:
\begin{equation}
Y(<r_{200}) d_{\rm A}(z)^2 = \frac{\sigma_{\rm T}}{m_{\rm e}c^2}\int_0^{r_{200}} 4\pi P_{\rm e}(r,z) r^2 dr, 
\end{equation}
where $d_{\rm A}$ is the angular diameter distance to the halo, $\sigma_{\rm T}$ is the Thomson cross-section, $m_{\rm e}$ the electron rest mass, and $P_{\rm e} = n_{\rm e}k_{\rm B}T_{\rm e}$ is the electron pressure with $k_{\rm B}$ being the Boltzmann constant. The flux therefore scales with the density of the hot gas, rather than the square of its density, as is the case for the collisional mechanisms that dominate the X-ray emissivity of diffuse plasmas. We compute $Y$ by summing the contributions of the gas particles associated with each halo, as per \citet{mccarthy17}. Star-forming gas particles (i.e. those comprising the ISM) are assumed to be neutral and do not contribute to the flux. 

The rows of Fig. \ref{fig:Ref100_Lx_tSZ} show the present-day $L_{\rm X}-M_{200}$ (top) and $Y d_{\rm A}^2 - M_{200}$ (bottom) relations\footnote{Since we are concerned with the response of $L_{\rm X}$ and $Y d_{\rm A}^2$ to deviations from the median scaling relations relating the properties of galaxies and their gaseous haloes to halo mass, precise correspondence between the properties of the simulated haloes and observational measurements is not crucial. Reasonable correspondence is however necessary to engender confidence in the realism of the simulations; we therefore compare the simulated scaling relations of $L_{\rm X}$ and $Y d_{\rm A}^2$ as a function of halo mass with observational measurements in Appendix \ref{app:obs}.}. Histograms of $L_{\rm X}$ and $Y d^2_{\rm A}$ are shown to the right of the main panel. In each panel the solid black line denotes the LOWESS running median. In the left-hand column symbols are coloured by their residuals with respect to the running median of the $f_{\rm CGM}-M_{200}$ relation. In both cases the colouring highlights that elevated values of the observable correspond to elevated values of $f_{\rm CGM}$. The running values of the Spearman rank correlation coefficient demonstrate that scatter about both proxies correlates strongly, significantly and positively with that about $f_{\rm CGM}$ for all $M_{200}$. The X-ray luminosity is somewhat noisier than the tSZ flux, which is unsurprising since it is also sensitive to the metallicity of the halo \citep[see e.g.][]{crain13}, and is more sensitive than the tSZ flux to the structure of the CGM.

The X-ray luminosity remains an attractive observable however, owing in particular to the dynamic range it displays: for haloes with $M_{200} \simeq 10^{12.5}\Msun$ the $10^{\rm th}-90^{\rm th}$ percentile range spans $1.54$ decades in $L_{\rm X}$. Measurements of the X-ray luminosity therefore afford the opportunity to highlight readily the diversity of halo properties at fixed $M_{200}$. We demonstrate this quantitatively by showing, as grey curves, the median $L_{\rm X}$ of the subsets of haloes representing the upper and lower quartiles of the $f_{\rm CGM}-M_{200}$ relation, in 10 bins of halo mass in the interval $10^{11.5} < M_{200} < 10^{13}\,\Msun$. The separation of the subsets peaks at $M_{200}=10^{12.2}\Msun$, with the gas-rich subset of haloes exhibiting a median $L_{\rm X}$ that is $1.5\,{\rm dex}$ greater than that of the gas-poor subset.

To use $L_{\rm X}$ and $Yd_{\rm A}^2$ to test the influence of BH mass on halo gas fractions, they must respond to scatter about the $M_{\rm BH}-M_{200}$ relation in a similar fashion to $f_{\rm CGM}$. In the centre column of Fig. \ref{fig:Ref100_Lx_tSZ}, the symbols are coloured by $\Delta \log_{10} M_{\rm BH}$. Similar to the top-right panel of Fig. \ref{fig:Ref100_megaplot}, residuals about the relations can be seen to correlate negatively with $\Delta \log_{10} M_{\rm BH}$. The running values of the Spearman rank correlation coefficient, shown below the main panels, behave similarly and become more strongly negative from relatively low masses to a peak at $M_{200} \simeq 10^{12.4}\Msun$. Therefore, the haloes of galaxies with central BHs that are more (less) massive than is typical for their mass are systematically `faint' (bright) in both $L_{\rm X}$ and $Yd_{\rm A}^2$. We again highlight the dynamic range of the X-ray luminosity and show as grey curves in the upper-centre panel the median $L_{\rm X}$ of the subsets representing the upper and lower quartiles of the $M_{\rm BH}-M_{200}$ relation. The subsets are again most strongly separated at $M_{200}=10^{12.2}\Msun$, with the subset of haloes with under-massive BHs exhibiting a median $L_{\rm X} = 1.2 \times 10^{40}\,{\rm erg s}^{-1}$, which is $1.4\,{\rm dex}$ greater than that of the over-massive BH subset, $L_{\rm X} = 4.5 \times 10^{38}\,{\rm erg s}^{-1}$.

At present dynamical measurements of the masses of BHs are available for only $\sim 10^2$ galaxies \citep[e.g.][]{kormendy13,mcconnell13}, presenting an obstacle to observational scrutiny of the influence of BH mass on halo gas fractions. However, as noted in Section \ref{sec:origin_of_scatter}, the correlations of $\Delta f_{\rm CGM}$ with $\Delta M_{\rm BH}$ and $\Delta \dot{M}_\star$ are of similar strength (but opposite sign). Identification of this correlation in observations would therefore corroborate EAGLE's predictions concerning the origin of scatter in the $f_{\rm CGM}-M_{200}$ scaling relation, hence it is important to establish how the proxies for $f_{\rm CGM}$ respond to scatter in SFR at fixed $M_{200}$. 

In the right-hand column of Fig. \ref{fig:Ref100_Lx_tSZ} the symbols are coloured by $\Delta \log_{10} \dot{M}_\star$. Encouragingly, we find that the residuals about the $\dot{M}_\star - M_{200}$ relation correlate positively with the residuals about both the $L_{\rm X} - M_{200}$ and $Yd_{\rm A}^2 - M_{200}$ relations, with the correlations being strong and significant for $M_{200} \lesssim 10^{12.7}\Msun$. Therefore, the haloes of galaxies that exhibit a SFR that is high (low) for their halo mass are systematically `bright' (faint) in both proxies. The grey curves of the upper-right panel show the median $L_{\rm X}$ of the subsets representing the upper and lower quartiles of the $\dot{M}_\star-M_{200}$ relation. The subsets are again most strongly separated at $M_{200}=10^{12.2}\Msun$, for which the upper quartile ($\dot{M}_\star > 1.8\,\Msunyr$) exhibit a median $L_{\rm X} = 1.8 \times 10^{40}\,{\rm erg s}^{-1}$, which is $1.7\,{\rm dex}$ (a factor of $\simeq 50$) greater than that of the lower quartile ($\dot{M}_\star < 0.2\,\Msunyr$), $L_{\rm X} = 3.8 \times 10^{38}\,{\rm erg s}^{-1}$.

\section{Summary and discussion}
\label{sec:summary}

We have examined the origin of scatter in the relationship between gas fraction, $f_{\rm CGM}$, and mass, $M_{200}$, of the haloes with mass similar to those that host present-day $\sim L^\star$ central galaxies ($10^{11.5} < M_{200} < 10^{13}\,\Msun$) in the EAGLE simulations. We quantify the scatter by computing the difference between each halo's gas fraction and the running median of the $f_{\rm CGM}-M_{200}$ relation, $\tilde{f}_{\rm CGM}(M_{200})$, constructed using the LOWESS locally-weighted scatterplot smoothing method.  

Our results are drawn primarily from the largest EAGLE simulation Ref-L100N1504, and its counterpart considering only collisionless gravitational dynamics, DMONLY-L100N1504. The parameters of the subgrid models governing feedback in the EAGLE Reference model were calibrated to ensure reproduction of key present-day properties of the galaxies, but the gaseous properties of galaxies and their haloes were not considered during the calibration and can be considered predictions of the simulations. We have also briefly studied the NOAGN-L050N0752 simulation in which AGN feedback is disabled, and its Reference model counterpart Ref-L050N0752.

Our findings can be summarized as follows:
\begin{enumerate}

\item Scatter about the $f_{\rm CGM} - M_{200}$ relation is not strongly correlated with residuals of the relationship between the stellar mass fraction of haloes and $M_{200}$. Low (high) halo gas fractions are therefore not generally a consequence of haloes having converted more (less) of their gas into stars throughout their assembly (Fig. \ref{fig:Ref100_mstar}).

\item Similarly, the scatter is neither strongly nor significantly correlated with the residuals of the relationship between the gas accretion rate of central BHs (whether measured instantaneously or time-averaged over 100 Myr) and $M_{200}$. Low (high) halo gas fractions are therefore not a consequence of relatively strong (weak) ongoing AGN feedback (Fig. \ref{fig:Ref100_megaplot}, top-left).

\item The scatter correlates strongly, significantly and negatively with the residuals of the relationship between the present-day mass of central BHs and $M_{200}$. At $M_{200}=10^{12.5}\Msun$ the Spearman rank correlation coefficient is $\rho^\prime = -0.75$. At fixed $M_{200}$ therefore, galaxies that host more-massive central BHs reside within relatively gas-poor haloes, and vice versa. This suggests that the main cause of scatter in $f_{\rm CGM}$ at fixed $M_{200}$ is differences in the mass of halo gas expelled by AGN feedback throughout the assembly of the halo (Fig. \ref{fig:Ref100_megaplot}, top-right).

\item A corollary of (ii) is the implication that the scatter about the $f_{\rm CGM} - M_{200}$ relation might be driven by a more fundamental process that fosters scatter in the $M_{\rm BH}-M_{200}$ relation. \citet{boothschaye10,boothschaye11} previously highlighted with cosmological simulations that this scatter is driven by differences in the binding energy of haloes at fixed mass. We find that scatter about the $f_{\rm CGM} - M_{200}$ relation indeed correlates strongly, significantly and negatively with the residuals of the $E_{\rm DMO}^{2500} - M_{200}$ relation, where $E_{\rm DMO}^{2500}$ is the intrinsic binding energy of the halo, i.e. that which emerges in the absence of the dissipative physics of galaxy formation, measured within $r_{2500}$. This correlation is strong and significant over a broad range in $M_{200}$, and at $M_{200}=10^{12.5}\Msun$ the correlation coefficient is $\rho^\prime = -0.62$. (Fig. \ref{fig:Ref100_megaplot}, bottom-left). 

\item Although reasonably interpreted as the fundamental origin of the scatter in $f_{\rm CGM}(M_{200})$, the influence of the intrinsic binding energy of haloes is communicated via AGN-driven gas expulsion for $M_{200}\gtrsim 10^{12}\Msun$. This is succinctly demonstrated by examination of the residuals about the $f_{\rm b}-M_{200}$ and $E_{\rm DMO}^{2500} - M_{200}$ relations (where $f_{\rm b}$ is the halo baryon fraction) in simulations with and without AGN feedback. In the Reference model simulation these residuals are strongly, significantly and negatively correlated for all $M_{200}$, whilst in the NOAGN model the correlation is weaker for relatively low-mass haloes ($M_{200} \lesssim 10^{12}\Msun$), and becomes positive for more massive haloes (Fig. \ref{fig:noagn}). 

\item The scatter in $f_{\rm CGM}(M_{200})$ correlates strongly, significantly and positively with the residuals of the relationship between the present-day SFR of central galaxies and $M_{200}$.  The correlation is similar to that with $\Delta M_{\rm BH}$, but with opposite sign. At $M_{200}=10^{12.5}\Msun$ the correlation coefficient is $\rho^\prime = 0.65$. Haloes with high (low) gas fractions for their mass therefore typically host central galaxies with high (low) SFRs. (Fig. \ref{fig:Ref100_megaplot}, bottom-right).

\item We consider the diffuse soft X-ray luminosity of the hot component of halo gas ($L_{\rm X}$) and the `flux' of the thermal Sunyaev-Zel'dovich effect ($Y d_{\rm A}^2$) as proxies for $f_{\rm CGM}$. Scatter about the relation of these observables with $M_{200}$ correlates positively with residuals of the $f_{\rm CGM}-M_{200}$ relation, such that variations in the halo gas fraction at fixed halo mass are echoed by the two observables. Residuals about the running median of both proxies also correlate negatively with scatter about the median BH mass at fixed $M_{200}$. At $M_{200}=10^{12.5}\Msun$ the associated correlation coefficients for $L_{\rm X}$ and $Yd_{\rm A}^2$ are $\rho^\prime = -0.62$ and $\rho^\prime = -0.51$, respectively. This highlights that they respond to variations in BH mass in a similar fashion to $f_{\rm CGM}$ (Fig. \ref{fig:Ref100_Lx_tSZ}, left and centre columns).

\item Scatter about the $L_{\rm X}-M_{200}$ and $Y d_{\rm A}^2-M_{200}$ relations also correlates strongly and positively with scatter about the median SFR at fixed $M_{200}$. At $M_{200}=10^{12.5}\Msun$ the associated correlation coefficients for $L_{\rm X}$ and $Yd_{\rm A}^2$ are $\rho^\prime = 0.68$ and $\rho^\prime = 0.43$, respectively. These correlations afford a route to observational scrutiny of the predictions of the simulations advanced here, without the need to acquire dynamical BH mass measurements for a large sample of galaxies. The simulations indicate that, for galaxies hosted by haloes of $M_{200} \simeq 10^{12.2}\Msun$, the median X-ray luminosity of those with $\dot{M}_\star > 1.8 \Msunyr$ (the upper quartile in SFR) is a factor $\simeq 50$ higher than for those comprising the lower quartile ($\dot{M}_\star < 0.2 \Msunyr$, Fig. \ref{fig:Ref100_Lx_tSZ}, right-hand column).
\end{enumerate}

The discovery of scaling relations connecting central BHs (with optical accretion discs of scale $\sim 10^{-2}\pc$) with the properties of galaxies \citep[on scales $\sim 10^3\pc$, e.g.][]{magorrian98,kormendy13} has focussed intense interest on the possibility of an intimate physical connection between the two. The release of rest mass energy from the accretion of gas onto BHs has long been advocated as a means to regulate cooling flows onto massive galaxies \citep[e.g.][]{silkrees98} at the centres of groups and clusters \citep[on scales $\sim 10^5-10^6\pc$, e.g.][]{binney95}. Cosmological simulations still lack the physics and resolution required to capture the full complexity of the coupling between these phenomena across such a broad dynamic range, but our findings nonetheless indicate that central BHs can also have a significant influence on the structure and content of the CGM. It is an unambiguous prediction of the EAGLE simulations that scatter in the central BH mass, at fixed halo mass, markedly influences the gas fractions of the haloes that host present-day $\sim L^\star$ central galaxies.

Our findings also suggest that it is possible to corroborate or falsify EAGLE's predictions for the origin of scatter about the $f_{\rm CGM}-M_{200}$ relation, using extant or forthcoming observations. We posit that the locally-brightest galaxy (LBG) sample from the New York University Value Added Galaxy Catalogue (Blanton et al. 2005, VAGC), based on the seventh data release of the Sloan Digital Sky Survey (SDSS/DR7, Abazajian
et al. 2009), is well-suited to this purpose. It has been used previously as the basis for stacked measurements of both diffuse X-ray luminosity \citep{anderson15} from the \textit{ROSAT} All-Sky Survey and the tSZ flux \citep{planck13,greco15} from \textit{Planck} maps. The acquisition of well-characterised rotation curves for this sample would further enable the identification of sub-samples with similar dynamical masses. We intend to pursue this line of enquiry in a forthcoming study; at present, X-ray luminosity remains the preferred proxy for $f_{\rm CGM}$, as the $\sim 10$ arcmin beam of \textit{Planck}'s 100 GHz maps corresponds to a scale comparable to or larger than $r_{200}$ for the majority of the LBG sample. However, characterisation of the tSZ flux within lower-mass haloes may soon be possible with the advent of next-generation, high-resolution, ground-based CMB experiments. 

A further, complementary means of assessing variations in $f_{\rm CGM}$ is to search for variations in the column densities of absorption systems, revealed by the intersection of bright quasars with the CGM of a sample of nearby galaxies. Such an approach would be similar in spirit, but different in detail, to the COS-AGN survey \citep{berg18}. Extending the COS-Halos survey \citep{tumlinson13}, COS-AGN enabled a controlled comparison of the absorption systems associated with galaxies with and without AGN. Consistent with our finding that the present-day BH accretion rate has little impact on halo gas fractions at fixed $M_{200}$, \citet{berg18} found no significant differences between these samples when examining the equivalent width distributions of absorption systems tracing the inner CGM. To probe the scenario advanced here, it is necessary instead to compare COS-Halos with a sample of galaxies with similar halo masses but diverse central BH masses (or a suitable proxy). Such a survey would also enable scrutiny of EAGLE's prediction that gas-rich (gas-poor) haloes exhibit higher (lower) SFRs than is typical for their halo mass, as a consequence of the negative correlation of scatter about both the $f_{\rm CGM}-M_{200}$ relation (due to the feedback history of the central BH) and the $\dot{M}_\star - M_{200}$ relation (due to the shift of the SF history to earlier times) with scatter about the $E_{\rm DMO}^{2500}-M_{200}$ relation. In a forthcoming study we will present expectations for the impact of BH mass and halo binding energy on the column densities and covering fractions of absorption species that are readily-accessible in the local Universe, such as C\textsc{iv}.

\section*{Acknowledgements}

JJD acknowledges an STFC doctoral studentship. RAC is a Royal Society University Research Fellow. BDO is supported by NASA ATP grant NNX16AB31G. MS is supported by the Netherlands Organisation for Scientific Research (NWO) through VENI grant 639.041.749. SM acknowledges support from the Academy of Finland, grant number 314238. This project has received funding from the European Research Council (ERC) under the European Union's Horizon 2020 research and innovation programme (grant agreement number 769130). The study made use of high performance computing facilities at Liverpool John Moores University, partly funded by the Royal Society and LJMU's Faculty of Engineering and Technology, and the DiRAC Data Centric system at Durham University, 
operated by the Institute for Computational Cosmology on behalf of the 
STFC DiRAC HPC Facility (www.dirac.ac.uk). This equipment was funded by 
BIS National E-infrastructure capital grant ST/K00042X/1, STFC capital 
grants ST/H008519/1 and ST/K00087X/1, STFC DiRAC Operations grant 
ST/K003267/1 and Durham University. DiRAC is part of the National 
E-Infrastructure.




\bibliographystyle{mnras}
\bibliography{bibliography} 




\appendix

\section{Comparison of simulated X-ray luminosities and tSZ fluxes with observations}
\label{app:obs}

\begin{figure}
\includegraphics[width=\columnwidth]{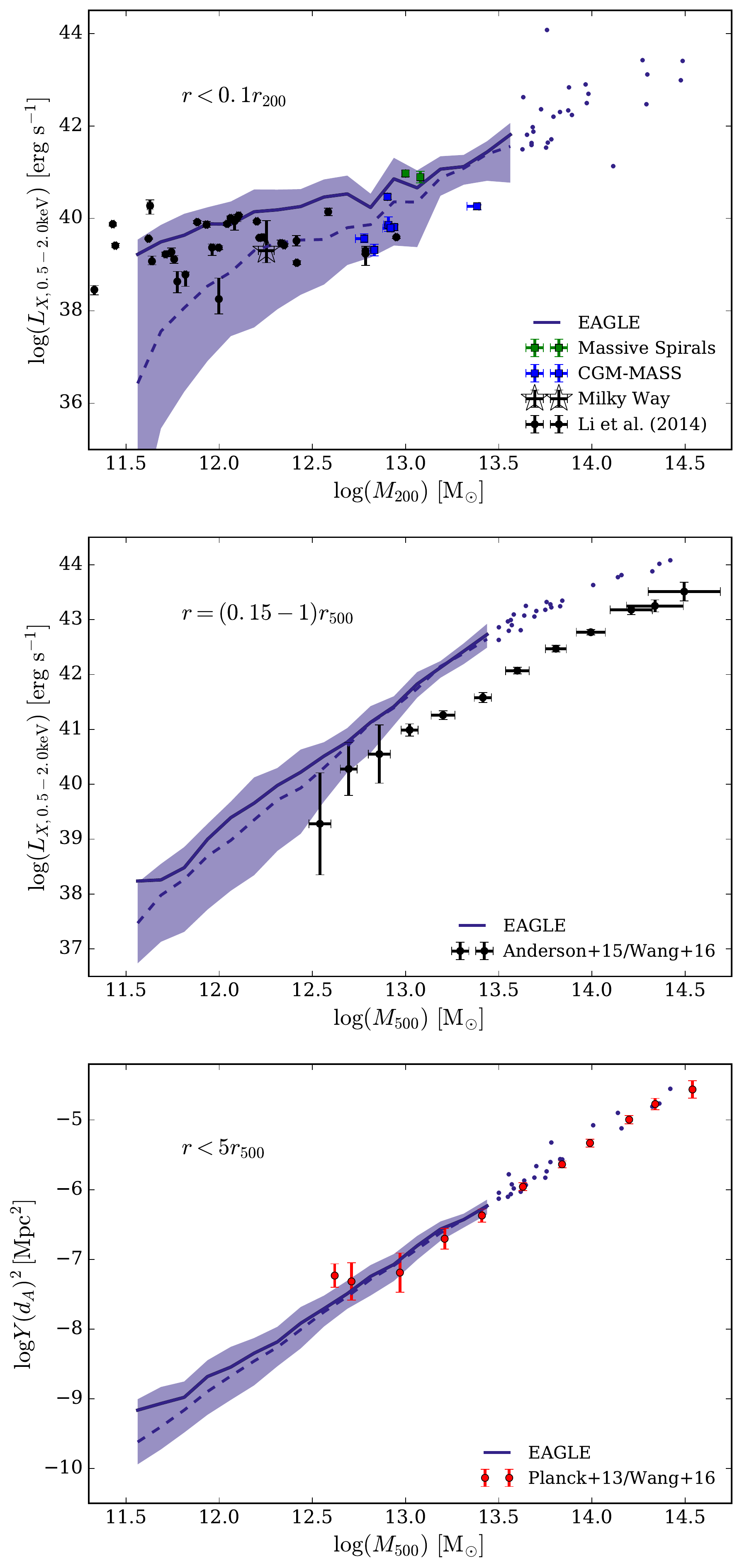} 
\caption{Comparison of EAGLE's CGM scaling relations with observational measurements. \textit{Top panel}: the $L_X-M_{200}$ relation, where $L_{\rm X}$ is computed within a spherical aperture of $r=0.1r_{200}$. Solid (dashed) lines indicating the binned mean (median) values, and shading indicates the $10^{\rm th}-90^{\rm th}$ percentile scatter. Galaxies in bins sampled by fewer than 10 galaxies are shown individually. Observational measurements are taken from the homogeneously-reanalysed compliation of \citet{li17}. \textit{Centre panel}: the $L_X-M_{500}$ relation, where $L_{\rm X}$ is computed within a spherical annulus of $0.15<r/r_{200}<1$. Observational measurements are based on X-ray and weak lensing shear maps stacked about the coordinates of massive local galaxies, as per \citet{wang16}. \textit{Bottom panel}: the $Yd_{\rm A}^2-M_{500}$ relation, where $Y$ is computed in a spherical aperture of $r=5r_{500}$. As above, observational measurements are  based on the stacking analysis of \citet{wang16}, here using \textit{Planck} CMB maps.}
\label{fig:observations}
\end{figure}

In this short Appendix we briefly compare the observable properties of haloes in the EAGLE simulation with observational measurements. In the panels of Fig. \ref{fig:observations}, we show the mean (solid lines) and median (dashed lines) values of scaling relations as a function of halo mass, with the $10^{\rm th}-90^{\rm th}$ percentile scatter shown via shading. In bins sampled by fewer than 10 galaxies, haloes are plotted individually. We remind the reader that, when comparing the simulations with measurements derived from stacked observational data, it is more appropriate to compare with the mean of the simulated sample rather than the median.

The upper panel of Fig. \ref{fig:observations} focusses on the $L_X-M_{200}$ relation. The observational measurements are comprised of \textit{XMM-Newton} and \textit{Chandra} direct detections of diffuse emission from the environments of massive, isolated spiral galaxies, compiled and homogeneously re-analysed by \citet{li17}. Specifically, these are observations of NGC 1961 \citep{bogdan13a,anderson16} and NGC 6753 \citep{bogdan13a}, labelled here as ``Massive Spirals", the CGM-MASS sample of \citet{li17}, a measurement of the Milky Way's X-ray luminosity from \citet{snowden97} with uncertainty estimates from \citet{miller15}, and a sample of inclined disc galaxies observed with \textit{Chandra} and presented by \citet{li14}. Since direct detections of diffuse X-ray emission from galaxies are generally limited to $r < 0.1r_{200}$, the luminosities of EAGLE haloes in this panel were computed using a 3-dimensional spherical aperture of this radius. The range of luminosities displayed by EAGLE galaxies is generally compatible with the scatter of the observational measurements, but we caution that direct detection observations are likely strongly biased towards the most X-ray-luminous galaxies at fixed mass. 

In the centre panel, we compare the X-ray luminosity of EAGLE galaxies to the $L_{\rm X}-M_{500}$ scaling relation presented by \cite{wang16}, who used weak lensing halo mass estimates to recalibrate the relation originally presented by \citet{anderson15}, derived by stacking X-ray maps from the \textit{ROSAT} All-Sky Survey about the coordinates of optically-selected massive galaxies in the local Universe. The low spatial resolution of the maps precludes the excision of bright X-ray point sources, therefore we compare to the "CGM" luminosities of \citet{anderson15}, computed using an aperture of $0.15 < r/r_{200}< 1$. This comparison reveals that the extended gas haloes of very massive galaxies, and galaxy groups, in EAGLE are too X-ray-luminous. This shortcoming was previously highlighted by \citet{schaye15}, who showed that the issue can be mitigated via the use of a higher temperature increment in EAGLE's stochastic heating implementation of AGN feedback. We elected to present results here from the Ref-L100N1504 simulation rather than the AGNdT9-L050N0752, which uses this higher heating temperature, since the latter affords a factor of 8 poorer sampling of the galaxy population. Moreover, we focus here primarily on $\sim L^\star$ galaxies, for which the correspondence with the observed X-ray luminosity is reasonable for the Reference model.

Finally, we show in the bottom panel the stacked tSZ signal for the same sample used by \citet{anderson15} and \citet{wang16}, as presented by \citet{planck13}. Following \citet{mccarthy17}, we convert the observed values of $Y(<r_{500})$ back into the the measured flux $Y(<5r_{500})$, and compare with tSZ flux of EAGLE galaxies computed within the same spherical aperture. The simulations reproduce the observed flux on these relatively large scales very well; a more stringent test of the simulation based on the tSZ flux awaits the availability of panoramic CMB maps with high spatial resolution.

We reiterate that the gaseous properties of galaxies and their haloes were not considered during the calibration of the parameters governing energetic feedback in EAGLE, and they may therefore be considered as predictions of the simulations. Whilst it is possible to identify differences in detail, the observable properties of the extended gaseous environments of $\sim L^{\star}$ galaxies in EAGLE are in sufficiently good agreement with observational measurements to engender confidence in this aspect of the simulations.



\bsp	
\label{lastpage}
\end{document}